\documentclass[twocolumn,aps,preprintnumbers,prl,showpacs,amsfonts,amsmath,amssymb,superscriptaddress,floatfix]{revtex4}
\usepackage[dvips]{graphicx}
\usepackage[]{verbatim}
\allowdisplaybreaks[3]
\begin{document}
\title{Controlled coupling of a single nitrogen-vacancy center to a silver nanowire}
\date{\today}
\author{Alexander Huck}\email{alexander.huck@fysik.dtu.dk}
\affiliation{Department of Physics, Technical University of Denmark, Building 309, 2800 Lyngby, Denmark}
\author{Shailesh Kumar}
\affiliation{Department of Physics, Technical University of Denmark, Building 309, 2800 Lyngby, Denmark}
\author{Abdul Shakoor}\thanks{Present address: School of Physics and Astronomy, University of St. Andrews, Fife KY16 9SS, UK}
\affiliation{Department of Physics, Technical University of Denmark, Building 309, 2800 Lyngby, Denmark}
\author{Ulrik L. Andersen}
\affiliation{Department of Physics, Technical University of Denmark, Building 309, 2800 Lyngby, Denmark}

\begin{abstract}
We report on the controlled coupling of a single nitrogen-vacancy (NV) center to a surface plasmon mode propagating
along a chemically grown silver nanowire. We locate and optically characterize a single NV-center in a uniform
dielectric environment before we controllably position this emitter in the close proximity of the nanowire. We are thus
able to control the coupling of this particular emitter to the nanowire and directly compare the photon emission
properties before and after the coupling. The excitation of single plasmonic modes is witnessed and a total rate
enhancement by a factor of up to 4.6 is demonstrated.
\end{abstract}

\pacs{73.20.Mf, 42.50.Ct} \maketitle

Strong enhancement of the fluorescence and scattered light from a single molecule, quantum dot, or nitrogen vacancy
(NV) center can be observed by placing the particle in the vicinity of a metallic
nano-sphere~\cite{2006Kuehn,2006Anger,2009Schietinger,1997Nie,1997Kneipp}. This enhancement arises due to the
excitation of strongly confined and localized surface plasmon on the metallic sphere. Despite the strong enhancement of
emissive processes, the resulting photon emission in those realizations does not couple preferably to a dedicated
spatial mode and are thus not directly suitable for applications in, for instance, quantum information processing where
the coupling between a single spatial mode and a single emitter is often a requirement for scalability.

To obtain an efficient coupling to a specific spatial mode while exploiting the enhancement properties of strongly
confined surface plasmons, it has been suggested to use propagating plasmons on cylindrical wires as opposed to
localized plasmons~\cite{2006Chang}. By using this method, it is possible to enhance the emissive process into one
particular propagating plasmon mode which can be further transferred into a photonic mode of an optical waveguide with
high efficiency~\cite{2008Pyayt,2009Chen}. Such a light-matter interface mediated by surface plasmons can be used to
efficiently generate single photons and to enable strong non-linearities at the single photon level which can be
exploited to make a single photon transistor~\cite{2007Chang} or to perform a near error-free deterministic Bell
measurement~\cite{2010Witthaut}.

The coupling of individual single photon emitters to propagating surface plasmon modes on individual silver nanowires
has been demonstrated for CdSe quantum dots~\cite{2007Akimov} and for NV-centers in nano-crystal
diamonds~\cite{2009Kolesov}. In these experiments, however, the wire-crystal systems were not deterministically
assembled thus rendering the rate enhancement estimation highly uncertain. By comparing an ensemble of uncoupled single
emitters with an ensemble of coupled single emitters, they estimated rate enhancement factors of 1.7~\cite{2007Akimov}
and 2.5~\cite{2009Kolesov}.

In this letter, we demonstrate the controlled coupling of a single NV-center in a diamond nano-crystal to a surface
plasmon mode propagating along a silver nanowire. The wire-crystal system is deterministically assembled by the use of
an atomic force microscope. This approach allows us to directly compare the emission properties of a particular
NV-center in a homogeneous dielectric environment with the emission properties of the same emitter but placed in the
near vicinity of a silver nanowire supporting a propagating surface plasmon mode. The coupling to the propagating
surface plasmon mode is evidenced by a decrease in the emitter lifetime and the emission of a single photon is verified
by a measurement of the second-order correlation function.

\begin{figure}[htbp]
\includegraphics[]{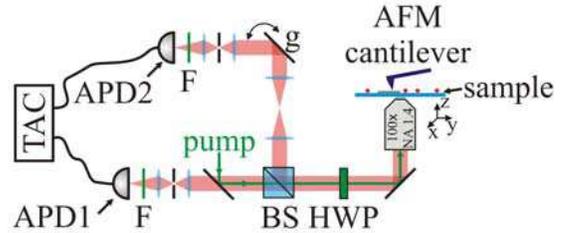}
\caption{(Color online). Experimental setup: TAC - time to amplitude converter, APD1 and APD2 - avalanche photo diode
detection channel 1 and 2, respectively, g - galvanometric mirror, BS - 50/50 beam splitter, and HWP - half waveplate.
The transmitted spectrum of the fluorescence filter F lies between 647 nm and 785 nm. \label{setup}}
\end{figure}

Our experimental setup is a home-built inverted scanning confocal microscope using an oil immersion objective with a
numerical aperture of 1.4 combined with a coaxially aligned atomic force microscope (AFM, NT-MDT SMENA), as illustrated
in Fig.~\ref{setup}. For the excitation of NV-centers we either use a continuous wave or a pulsed laser, both operating
at a vacuum wavelength of 532~nm. The pulsed laser has a repetition rate of 5.05~MHz and a pulse width of 4.6~ps. The
fluorescence emission from the sample is collected by two optical channels and detected with avalanche photo diodes
APD1 and APD2, respectively. The channel of APD1 is directly aligned with the pump beam, while the channel of APD2 is
decoupled from the pump beam via a galvanometric mirror, as illustrated in Fig.~\ref{setup}. The fluorescence filters
placed in front of both detectors are highly transmissive for photon wavelength between 647~nm and 785~nm. APD1 and
APD2 compose, together with the beam splitter BS, the well known Hanbury-Brown and Twiss interferometer used for the
characterization of a single photon source~\cite{1956HBT}. The AFM can be operated in tapping mode or contact mode.
Tapping mode operation is used to record the topography of the sample and with this to measure the height of the
nanowire and the diamond crystals as well as the lateral position of individual nano-crystals with respect to the
silver nanowire. By using the AFM in contact mode operation we are able to controllably push a chosen diamond
nano-crystal across the sample. This technique is used to place a single diamond nano-crystal near a silver nanowire,
thus controllably coupling the NV-center to the propagating plasmonic mode.

\begin{figure}[htbp]
\includegraphics[]{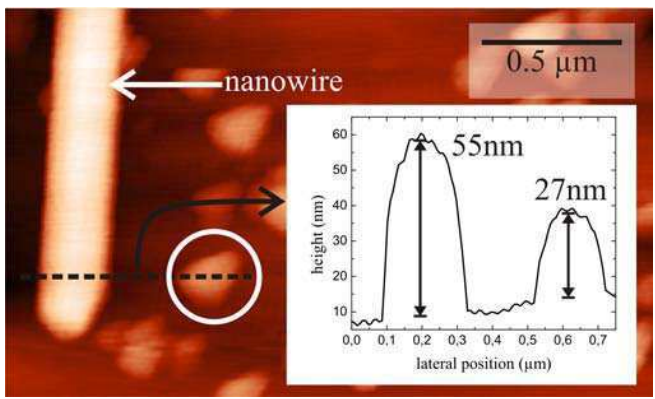}
\caption{(Color online). AFM topography of the nanowire and the nano-diamond containing the investigated NV-center
(white circle). The inset shows the height profile of the wire and the diamond, taken along the black dashed line.
\label{AFM_initial}}
\end{figure}

The sample was prepared on a plasma cleaned fused silica substrate. After the cleaning process, we first spin-coat
nano-diamonds with sizes $< 50 \text{nm}$ (MSY 0-0.05, Microdiamt AG) on the sample. Only a few of the nano-diamonds
are optically active containing a single NV-defect. Chemically grown silver nanowires with diameters in the range $50 -
70 \text{nm}$ and length of $1 - 10 \mu\text{m}$ fabricated by a polyol reduction process of silver
nitrate~\cite{2008Korte} were subsequently spin-coated on the substrate resulting in a sufficient number of individual
and well separated nanowires.

\begin{figure}[htbp]
\includegraphics[width=0.48\textwidth]{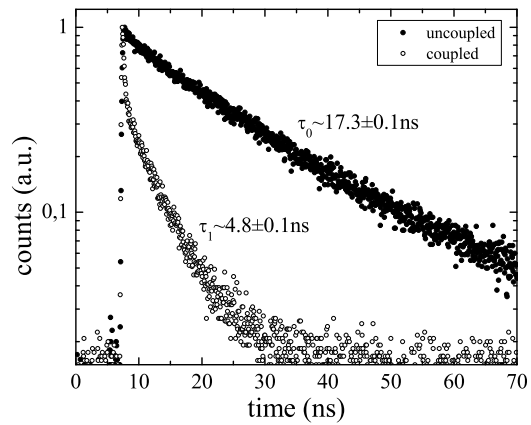}
\caption{Fluorescence lifetime measurement of the uncoupled NV-center in a uniform dielectric environment (black dots)
and of the same NV-center after it has been coupled to the propagating SPP mode of the silver nanowire (open circles).
\label{figure_lifetime}}
\end{figure}

\begin{figure}[htbp]
\includegraphics[width=0.48\textwidth]{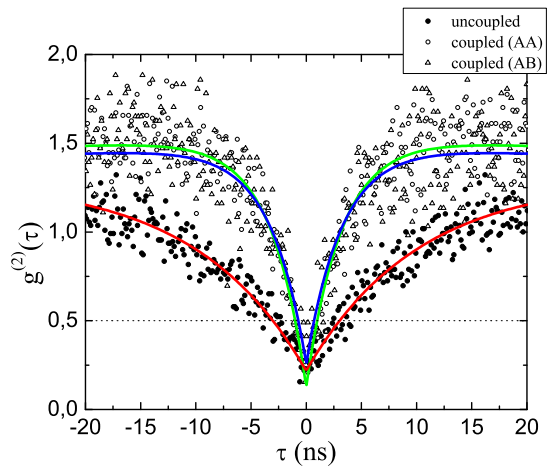}
\caption{(Color online). Second order correlation function measurement of the uncoupled NV-center (black dots) and the
coupled NV-center with both detectors aligned to spot 'A' (open dots) and APD1 aligned to spot 'A' and APD2 aligned to
spot 'B' (open triangles). The lines are a best fit to the data: red - uncoupled, green - coupled AA, and blue -
coupled AB. \label{figure_correlation}}
\end{figure}

We started the experiment by recording a fluorescence image with APD1. The AFM topography image of the same area was
then used to locate a suitable nano-diamond containing a single NV-defect. This diamond was separated from other
diamonds and brought near the nanowire for characterizing the NV-center in a uniform dielectric environment. At this
step of the experiment, the diamond was still well separated from the silver nanowire, as illustrated by the
AFM-topography image shown in Fig.~\ref{AFM_initial}. In the inset of Fig.~\ref{AFM_initial} we present the height
profile of the silver nanowire and the diamond, which at maximum were measured to be $55$~nm and $27$~nm, respectively.
The length of the nanowire was measured with the AFM to be about $2.8\mu\text{m}$. With the diamond in this position,
we measure the decay time $\tau_0$ of the NV-center's excited state, which is shown by the black dots in
Fig.~\ref{figure_lifetime}. A single exponential fit yields an excited state lifetime of $\tau_0 \approx 17.3 \pm
0.1$~ns. In contrast to NV-centers in bulk diamond, where an excited state lifetime of $\tau_{0,bulk} \approx 12 \,
\text{ns}$ is usually measured~\cite{1983Collins,2008Batalov}, we attribute the increased lifetime to the decreased
effective refractive index of the surroundings, which is significantly smaller than the refractive index of bulk
diamond. The measured second order correlation function $g^{2}(\tau)$ of this NV-center is presented by the black dots
in Fig.~\ref{figure_correlation} and the red line is a best fit to the data using the model of
Ref.~\cite{2000Kurtsiefer}. The emission of single photons is clearly confirmed by the fact that $g^{(2)}(0)$ is well
below $0.5$. By measuring the fluorescence count rate as a function of the linear polarization of the pump beam, we
verified that a large component of the NV-centers dipole moment is aligned perpendicular to the nanowire
axis~\cite{2005Epstein, 2009Schietinger}.

\begin{figure}[htbp]
\includegraphics[]{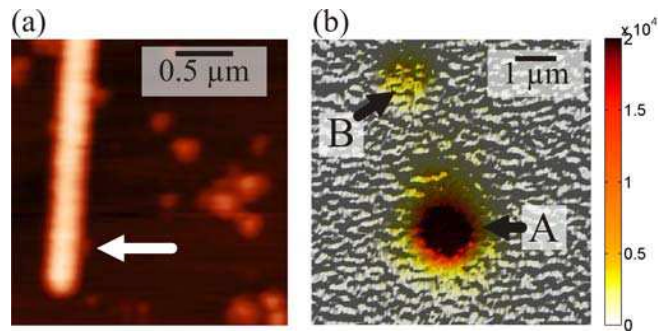}
\caption{(Color online). (a) AFM image taken after the nano-diamond has been located near the wire. The location of the
diamond is indicated by the white arrow. (b) Photoluminescence image of the coupled NV-center nanowire system, taken
while continuously exciting the NV-center, scanning the sample with the galvanometric mirror and recording the signal
with APD2. Encoded in the color scale is the APD signal in counts/s. \label{AFM_coupled_galvo}}
\end{figure}

As a next step, we carefully push the nano-diamond in the very close proximity of the nanowire by operating the AFM in
contact mode. The final position of the nano-diamond near the silver nanowire is indicated by the white arrow in the
AFM topography image shown in Fig.~\ref{AFM_coupled_galvo} (a). The NV-centers dipole moment is still largely aligned
perpendicular to the nanowire axis, which was verified by a new measurement of the count rate as a function of the
linear polarization of the pump beam. As a consequence, the NV-center now feels a changed dielectric environment
compared to its previous position. This is witnessed by the decrease of its excited state lifetime, as shown by the
open circles in Fig.~\ref{figure_lifetime}. Directly compared to the lifetime $\tau_0$ of the NV-center in its previous
position, we measured a decrease of the excited states lifetime of the coupled system $\tau_1$ by a factor of
$\tau_0/\tau_1 \approx 3.6 \pm 0.1.$ Now, the NV-center does not only radiate to the far field, but also couples to the
propagating plasmonic mode of the silver nanowire~\cite{2006Chang}.

Evidence for the excitation of the propagating plasmonic mode is given by the fluorescence image recorded with the
detector APD2, which is shown in Fig.~\ref{AFM_coupled_galvo} (b). This image has been obtained by continuously
exciting the NV-center while scanning the image plane using the galvanometric mirror. Two emission spots can be seen in
Fig.~\ref{AFM_coupled_galvo} (b). The emission spot labeled as 'A' in Fig.~\ref{AFM_coupled_galvo} (b) comprises the
radiative emission from the NV-center together with emission from the nearby nanowire end face. Spot 'B' in
Fig.~\ref{AFM_coupled_galvo} (b) only comprises emission from the far nanowire end face. The intensity measured from
spot 'B' is thus proportional to the coupling of the NV-center to the propagating plasmonic
mode~\cite{2006Chang,2007Akimov,2009Kolesov}. However, an exact estimation of the NV-center coupling to the propagating
plasmonic mode by measuring the intensity of spot 'B' is difficult due to plasmon propagation losses along the nanowire
and the complex reflections at the nanowire end face in conjunction with the broad emission spectrum of the NV-center.

The $g^{(2)}(\tau)$ function measurement with the detectors APD1 and APD2 being both aligned on emission spot 'A' is
shown by the open dots in Fig.~\ref{figure_correlation}. With the open triangles we present the $g^{(2)}(\tau)$
function measurement between emission spot 'A' and 'B'. The green and blue line represent a best fit to the data for
both measurement realizations. In both cases, a $g^{(2)}(0) < 0.5$ confirms the generation of single photons, and the
excitation of \emph{single} propagating surface plasmon modes is confirmed by the measurement between spot 'A' and 'B'.

\begin{figure}[h!]
\includegraphics[width=0.49\textwidth]{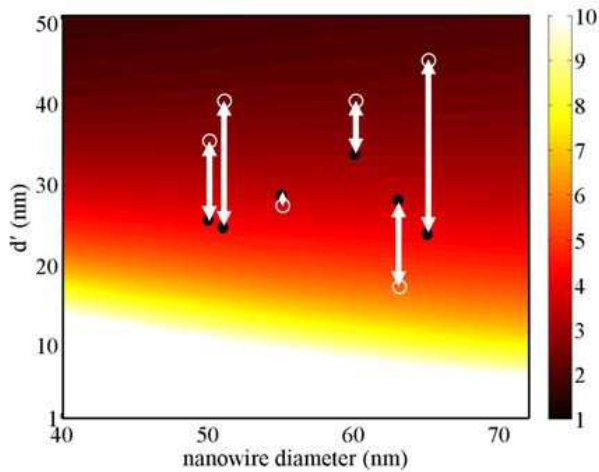}
\caption{(Color online). Total rate enhancement as a function of nanowire diameter and NV-center distance from the
nanowire surface $d^{\prime}.$ The open white circles indicate the minimum expected total rate enhancement for each
nanowire diameter and diamond height, the black filled dots are the experimentally measured rate enhancement factors
for each nanowire diameter, and the white arrows are a guide to the eye to indicate data related via the same nanowire
diameter. \label{theory}}
\end{figure}

Finally, we compare our experimental results with theoretical predictions~\cite{2006Chang}. In Fig.~\ref{theory} we
plot the total rate enhancement for the relevant range of nanowire diameter and emitter distance from the nanowire
surface. The calculation is done for a vacuum wavelength of $700~\text{nm}$ and we use the Drude model to estimate the
electric permittivity $\epsilon_{Ag}$ of silver~\cite{1972Johnsen}. In order to account for the relatively high
refractive index of diamond, the electric permittivity of the dielectric medium $\epsilon_1$ surrounding the nanowire
was set to $3$ in the calculation. For the nanowire diameter of $55~\text{nm}$ and a diamond height of $27~\text{nm}$
we expect from these calculations a total rate enhancement of $3.8$, assuming that the NV-center is located at the
maximum position of $27~\text{nm}$ away from the nanowire surface and that the NV-centers dipole moment is aligned
parallel to the radial electric field component $\mathbf{E}_r$ of the propagating plasmonic mode. This expected rate
enhancement is in good agreement with our experimental result of $3.6\pm0.1$. We emphasize that only for parallel
alignment of the NV-centers dipole moment to $\mathbf{E}_r$, an efficient coupling to the propagating plasmonic mode is
achieved, since for the relevant range of parameter $|\mathbf{E}_z| \ll |\mathbf{E}_r|,$ where $\mathbf{E}_z$ is the
plasmon electric field along the nanowire axis.

In TABLE~\ref{table_summary_coupling} we summarize the relevant physical parameters for various NV-center/nanowire
systems that we have successfully assembled. For all realizations we observe the excitation of a single surface
plasmon, confirmed by $g^{(2)}(0)<0.5$. The coupling to the propagating plasmonic mode was witnessed by strong rate
enhancement factors $\tau_0/\tau_1$ and verified by the observation of photon emission from the nanowire end face. In
addition to the experimentally obtained rate enhancements we also present the minimum expected rate enhancement in
column four of TABLE~\ref{table_summary_coupling}, calculated under the assumption that the NV-centers dipole moment is
aligned parallel to $\mathbf{E}_r$. All measured and calculated enhancement factors are graphically illustrated in
Fig.~\ref{theory}.
\begin{table}[htbp]
\begin{tabular}{c|c|c|c}
  nanowire     & diamond     & \textit{total rate}       & min. \\
  diameter     & height      & \textit{enhancement}      & expected rate \\
   (nm)        & (nm)        & \textit{$\tau_0/\tau_1$} & enhancement \\  \hline \hline
  65 & 45 & $\textit{4.6$\pm$0.1}$ & 2.2 \\
  60 & 40 & $\textit{2.9$\pm$0.1}$ & 2.4 \\
  51 & 40 & $\textit{4.4$\pm$0.2}$ & 2.4 \\
  63 & 17 & $\textit{3.6$\pm$0.1}$ & 5.7 \\
  50 & 35 & $\textit{4.2$\pm$0.1}$ & 2.8 \\
\end{tabular}
\caption{Summary of the relevant physical parameters of successfully assembled NV-center/nanowire systems.
\label{table_summary_coupling}}
\end{table}

In view of improving the coupling efficiency of the NV-center to the plasmonic mode it is necessary to decrease both
the diameter of the nanowire and the size of the diamond, as can be seen from Fig.~\ref{theory}. Thinner nanowires can,
for instance, be obtained by an optimization of the nanowire fabrication process~\cite{2008Korte}. As an alternative,
it might also be possible to sculpture nanowires from chemically prepared metallic flakes using focused ion beam
milling~\cite{2010Huang}. The latter technique will have the advantage of gaining ultimate control over the nanowire
dimensions. We have found experimentally, that nanowires made by electron beam lithography and thermal metal deposition
are not suitable for coupling single NV-centers to their propagating plasmonic mode. Fluorescence from those nanowires
largely overlaps with the emission spectrum of a NV-center, which as a direct consequence limits the possibility of
detecting single photons. Surface roughness of lithographically prepared nanowires further limits the propagation
distance of plasmonic modes~\cite{2005Ditlbacher}. Nano-diamonds with sub-10~nm diameters containing single NV-centers
with stable photon emission rates have recently been reported in the literature~\cite{2009Tisler,2010Bradac}.

In conclusion, by the aid of an atomic force microscope we have nano-assembled a system comprising a single NV center
in a nano-crystal diamond and a chemically grown silver nanowire. This method allowed us to directly compared the
emission properties of a single NV-center in a uniform dielectric environment with the emission properties of the same
emitter coupled to the nano-wire. An enhancement of the NV-centers decay rate by a factor of 4.6 is directly measured
and the excitation of single surface plasmons is evidenced by the observation of single photon emission at the far end
of the nanowire. We believe that the method presented in this article in combination with smaller diamonds and
optimized metallic structures will pave the way for strongly coupled plasmonic systems.

We gratefully acknowledge fruitful discussions with Anders S. S\o rensen and Fedor Jelezko. This project has been
financially supported by the Villum Kann Rasmussen foundation, the Carlsberg foundation, and the Danish council for
independent research - natural sciences (FNU).


\begin{thebibliography}{99}
\bibitem{2006Kuehn} S. K{\"u}hn, U. H{\aa}kanson, L. Rogobete, and V. Sandoghdar, Phys. Rev. Lett \textbf{97}, 017402 (2006).

\bibitem{2006Anger} P. Anger, P. Bharadwaj, and L. Novotny, Phys. Rev. Lett. \textbf{96}, 113002 (2006).

\bibitem{2009Schietinger} S. Schietinger et al., Nano Lett.\textbf{9}, 1694-1698 (2009).

\bibitem{1997Nie} S. Nie and S. R. Emory, Science \textbf{275}, 1102--1106 (1997).

\bibitem{1997Kneipp} K. Kneipp et al., Phys. Rev. Lett. \textbf{78}, 1667--1670 (1997).

\bibitem{2006Chang} D.E. Chang, A.S. S\o rensen, P.R. Hemmer, and M.D. Lukin, Phys. Rev. Lett. \textbf{97}, 053002 (2006), D.E. Chang, A.S. S\o rensen, P.R. Hemmer, and M.D. Lukin, Phys. Rev. B \textbf{76}, 035420 (2007).

\bibitem{2008Pyayt} A. L. Pyayt et al., Nature Nanotechnology \textbf{3}, 660--665 (2008).

\bibitem{2009Chen} X.-W. Chen et. al., Nano Lett. \textbf{9}, 3756--3761 (2009).

\bibitem{2007Chang} D.E. Chang, A.S. S\o rensen, E.A. Demler, and M.D. Lukin, Nature Physics (London) \textbf{3}, 807 (2007).

\bibitem{2010Witthaut} D. Witthaut et. al., arXiv:1007.3273 (2010).

\bibitem{2007Akimov} A.V. Akimov et al., Nature (London) \textbf{450}, 402 (2007).

\bibitem{2009Kolesov} R. Kolesov et al., Nature Physics \textbf{5}, 470-474 (2009).

\bibitem{1956HBT} R. Hanbury Brown and R. Q. Twiss, Nature \textbf{178}, 1046--1048 (1956).


\bibitem{1983Collins} A. T. Collins et al., J. Phys. C, Solid State Phys. \textbf{16} (11), 2177–2181 (1983).

\bibitem{2008Batalov} A. Batalov et. al., Phys. Rev. Lett \textbf{100}, 077401 (2008).

\bibitem{2000Kurtsiefer} C. Kurtsiefer, Sonja Mayer, Patrick Zarda, and Harald Weinfurter, Phys. Rev. Lett. \textbf{85}, 290–-293 (2000).

\bibitem{2005Epstein} R.J. Epstein et. al., Nature Physics \textbf{1}, 94--98 (2005).

\bibitem{1972Johnsen} P.B. Johnsen and R.W. Christy, PRB \textbf{6}, 4370 (1972).


\bibitem{2008Korte} Kylee E. Korte, Sara E. Skrabalak, and Younan Xia. Journal of Materials Chemistry \textbf{18}, 437–441 (2008).

\bibitem{2010Huang} J.-S. Huang et. al., arXiv:1004.1961 (2010).

\bibitem{2005Ditlbacher} H. Ditlbacher et. al., Phys. Rev. Lett. \textbf{95}, 257403 (2005).

\bibitem{2009Tisler} J. Tisler et al., ACS Nano \textbf{3}, 1959--1965 (2009).

\bibitem{2010Bradac} C. Bradac et al., Nature Nanotechnology \textbf{5}, 345--349 (2010).

\end{thebibliography}
\end{document}